\documentclass[journal=langd5,manuscript=article]{achemso}
\setkeys{acs}{keywords = true}
\setkeys{acs}{usetitle = true}
\usepackage[version=3]{mhchem} 
\usepackage{color}
\usepackage{threeparttable}

\usepackage{lineno}
\usepackage{hyperref}
\usepackage{upgreek}
\usepackage{setspace}

\author{Jiawang Cui}
\affiliation{State Key Laboratory of Engines, Tianjin University, Tianjin, 300350, China.}
\author{Tianyou Wang}
\affiliation{State Key Laboratory of Engines, Tianjin University, Tianjin, 300350, China.}
\alsoaffiliation{National Industry-Education Platform of Energy Storage, Tianjin University, Tianjin, 300350, China.}
\author{Zhizhao Che}
\email{chezhizhao@tju.edu.cn}
\affiliation{State Key Laboratory of Engines, Tianjin University, Tianjin, 300350, China.}
\alsoaffiliation{National Industry-Education Platform of Energy Storage, Tianjin University, Tianjin, 300350, China.}

\title{Freezing-melting mediated dewetting transition for droplets on superhydrophobic surfaces with condensation}

\SectionNumbersOn
\begin{document}
\setstretch{1.2}
\begin{abstract}
The water-repellence properties of superhydrophobic surfaces make them promising for many applications. However, in some extreme environments, such as high humidities and low temperatures, condensation on the surface is inevitable, which induces the loss of surface superhydrophobicity. In this study, we propose a freezing-melting strategy to achieve the dewetting transition from the Wenzel state to the Cassie-Baxter state. It requires freezing the droplet by reducing the substrate temperature and then melting the droplet by heating the substrate. The condensation-induced wetting transition from the Cassie-Baxter state to the Wenzel state is analyzed first. Two kinds of superhydrophobic surfaces, i.e., single-scale nano-structured superhydrophobic surface and hierarchical-scale micro-nano-structured superhydrophobic surface, are compared and their effects on the static contact states and impact processes of droplets are analyzed. The mechanism for the dewetting transition is analyzed by exploring the differences in the micro/nano-structures of the surfaces and it is attributed to the unique structure and strength of the superhydrophobic surface. These findings will enrich our understanding of the droplet-surface interaction involving phase changes and have great application prospects for the design of superhydrophobic surfaces.
\end{abstract}

\section{Introduction}\label{sec:1}
Superhydrophobic surfaces have numerous applications such as self-cleaning, anti-corrosion, anti-fogging, anti-icing, flow drag reduction, and heat transfer enhancement \cite{feng22, jeevahan18, liravi20, vazirinasab18, zhang15, Chu19APL, Chu24AM}. They are designed by inducing the Cassie-Baxter states of droplets, where air pockets exist beneath the droplet, preventing close contact of the droplet with the surface. However, the air pockets are gone when condensate droplets invade and fill the cavities between the micro/nano-structures of superhydrophobic surfaces, leading to the Wenzle states of the droplets, and hence the loss of superhydrophobicity. Therefore, the regulation of droplet contact states on superhydrophobic surfaces, especially in terms of the wetting and dewetting transitions, is important in both fundamental research and practical applications.

Many factors can lead to the wetting transition of a droplet from a Cassie-Baxter state to a Wenzel state, such as high-speed impact \cite{lambley20,shi19, lee15, ryu17}, mechanical squeeze \cite{fang18, garciagonzalez22, li17}, mechanical vibration \cite{boreyko09, lei14, wu16}, electric field \cite{barberoglou10, mchale19, mugele05, roy18}, magnetic field \cite{alazawi17, cheng12, zhou11}, and condensation \cite{furuta10, shen21, wen17, zhao18}. It is worth noting that condensation is the most common cause of wetting transition because it is prone to occur as long as the ambient temperature is below the dew-point temperature. For example, Furuta et al. \cite{furuta10} studied the influence of condensation on the wettability of rough and smooth superhydrophobic coatings. The result indicated that although there are differences in condensation conditions between these two surfaces, their contact angles exhibit strong temperature dependence at the dew point because of the decreasing air amount at the solid-liquid interface. In another example, Zhao et al. \cite{zhao18} reported wetting stabilities of two superhydrophobic surfaces with different nano-asperities packing densities under the effect of condensation. The result showed that when two small condensate droplets in the Cassie state merge into a larger droplet, the contact line stays unchanged on one surface while contracting on the other surface due to the different degrees of counter-lock of the solid/water interface. In general, the condensation characteristic on superhydrophobic surfaces and its influence on the wetting states of droplets are important and worthy of in-depth research.

As for the dewetting transition of droplets from a Wenzel state to a Cassie-Baxter state, it is difficult because the Wenzel state is generally energetically favorable on most surfaces \cite{verho12}. Some methods have been proposed to achieve the dewetting transition, such as droplet self-propelled motion due to surface deformation \cite{wen17ACSAMI, zhao20, boreyko09prl, zhang16}, mechanical vibration \cite{boreyko09, lei14, wu16, zhang19}, electric field \cite{barberoglou10, kim10, manukyan11}, magnetic field \cite{cheng12}, heating \cite{liu11}, internal flow \cite{Wang2022spontaneous}, and nonuniform freezing \cite{graeber17, lambley23}. For example, Wen et al. \cite{wen17ACSAMI} presented a strategy to achieve the dewetting transition by using a superhydrophobic surface to manipulate the initial position of condensate droplets by spacing nanowires closely and then to promote the spontaneous outward movement of these droplets from the structural gap by using microwire arrays. In another example, Lambley et al. \cite{lambley23} inhibited the degradation of droplet contact state by inducing droplet freezing on superhydrophobic surfaces across low ambient temperatures and pressures. Due to the local explosive vaporization during the nucleation, the freezing process was asymmetric and then may make the droplet expel from the surface instead of immersing into the surface structures.

It should be noted that these methods for achieving the dewetting transition are limited due to their strict implementation requirements, such as the high heating temperature \cite{liu11, wang14}, the utilization of magnetic fluid \cite{alazawi17, cheng12, zhou11}, the intrusion into the droplet \cite{barberoglou10, manukyan11}, the vacuum environment \cite{graeber17, lambley23, verho12}, or the complex structural design \cite{boreyko13, cunjing15}. Here, we propose a freezing-melting strategy to achieve the dewetting transitions on a superhydrophobic surface, which is a kind of single-scale nano-structured superhydrophobic surface. A superhydrophobic surface with similar chemical compositions but different structures (i.e., hierarchical-scale micro-nano-structures) was used for comparison. The wetting transition due to the condensation at reduced temperatures, which is set as the initial condition before the dewetting transfer, is also considered. The effects of the condensation on the static contact states and the dynamic impact processes of droplets on the two superhydrophobic surfaces are explored and compared. During the dewetting transitions under different temperature conditions, the shapes of droplets are measured, and the contact diameter and contact angle are analyzed. The mechanism of achieving the dewetting transition is investigated by substrate surface characterization before and after the dewetting, and the requirement for the superhydrophobic surface to achieve the dewetting transition is analyzed.

\section{Experimental section}\label{sec:2}
\begin{figure}
  \centering
  \includegraphics[scale=0.65]{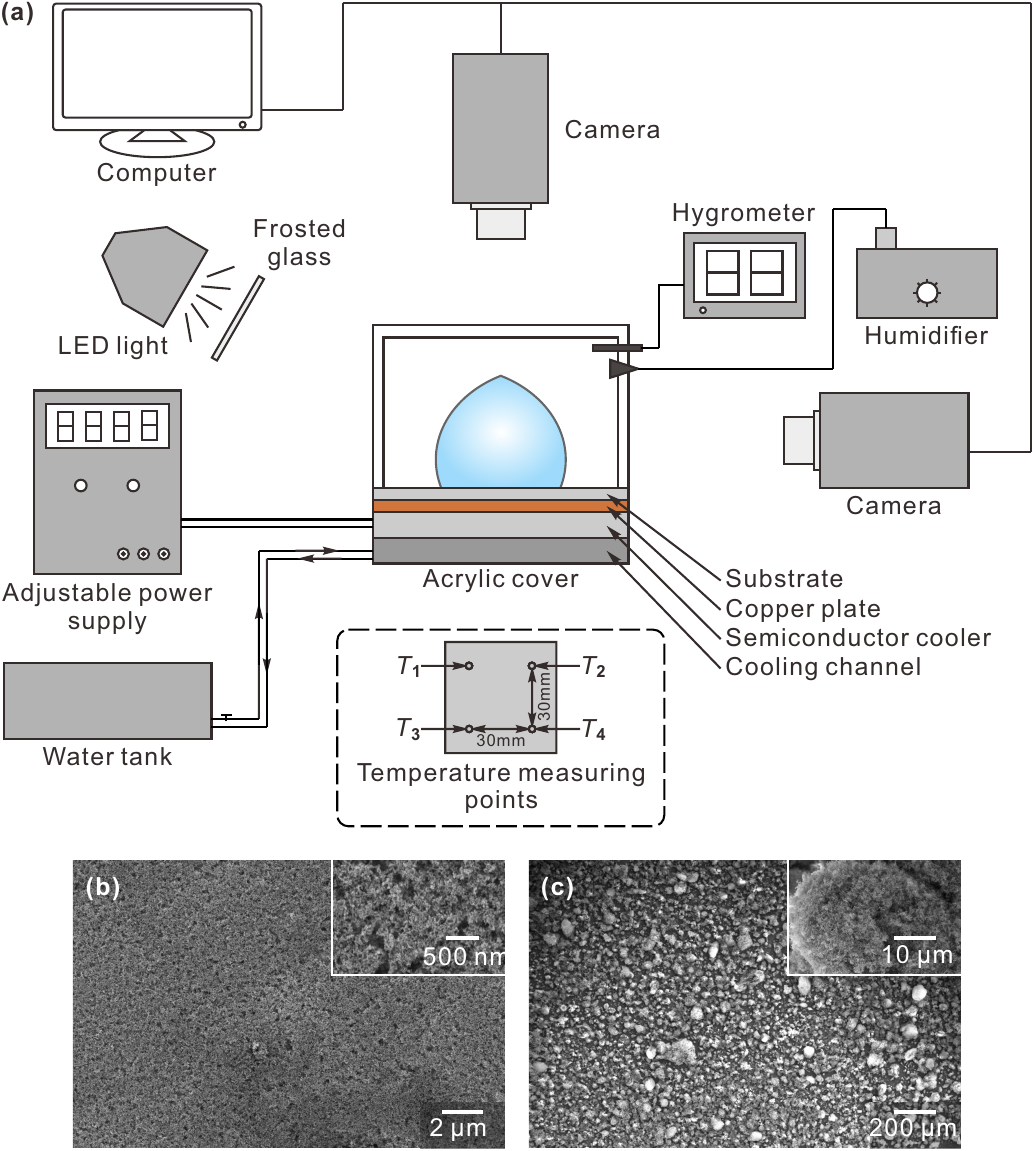}
  \caption{(a) Schematic diagram of the experimental setup. (b, c) SEM images of the two superhydrophobic surfaces: (b) single-scale nano-structured (SN) superhydrophobic surface; (c) hierarchical-scale micro-nano-structured (HMN) superhydrophobic surface.}\label{fig:01}
\end{figure}

The experimental setup is similar to that of our previous study \cite{cui23melting}, as shown in Figure \ref{fig:01}(a). The temperature of the superhydrophobic substrate was controlled by a semiconductor temperature-control system. An acrylic cover ($50 \times 50 \times 50$ cm$^3$) was used to maintain a stable environment for the condensation, freezing, and melting processes. The humidity of the operating condition inside the acrylic cover was maintained at 60\% by a humidifier and monitored by a hygrometer. Before the experiment, the humidifier was turned on to increase the humidity inside the acrylic cover until the humidity reached 60\%. For the study of the condensation under dew-point temperature (14.8--16.8 $^\circ$C at ambient temperature of 23--25 $^\circ$C and humidity of 60\%), different superhydrophobic surfaces were then placed on the copper plate, which had been pre-cooled to the specified temperature (i.e., the substrate temperature). The pre-cooling of the substrate, instead of reducing the substrate temperature after the deposition of the droplet, allows better control accuracy of the substrate temperature. Then, the condensation on the substrate was recorded by using a high-resolution CMOS camera (FLIR BFS-U3-17S7M-C) with a long-distance microscope (NAVITAR Zoom 6000) from the top view. For the study on the time variation of the contact angles, different superhydrophobic surfaces were set on the cooled copper plate as described above, and after a certain time (i.e., the cooling time mentioned in the later part of the paper), a droplet was deposited on the superhydrophobic surface. The static contact angle was recorded by using a high-resolution CMOS camera (FLIR BFS-U3-17S7M-C) with a macro lens (TOKINA AT-X PRO 100mm) from the side view. The effect of condensation on the droplet impact process was recorded by a high-speed camera (Photron SA 1.1) with a macro lens (TOKINA AT-X PRO 100mm) from the side view. After the morphology of the droplet stabilized, the temperature of the copper plate was changed by adjusting the power supply. Due to the quick temperature response of the semiconductor temperature-control system, the temperature of the copper plate can rapidly change (e.g., the time for the copper plate to decrease from 2 $^\circ$C to -25 $^\circ$C was less than 15 s). Under different conditions of temperature variation, the droplet on the superhydrophobic surface may wet, freeze, or melt. The change of droplet morphology was recorded by a high-resolution CMOS camera (FLIR BFS-U3-17S7M-C) with a macro lens (TOKINA AT-X PRO 100mm) from the side view.

Two kinds of superhydrophobic surfaces were prepared by creating micro/nano-structures on silicon substrates \cite{cui23melting}. One was prepared by spraying a mixed solution (85\% isopropanol, 12\% liquefied petroleum gas, and 3\% hydrophobic silica powder) on a silicon wafer. After drying, silica particles were stacked on the substrate surface and the surface achieved a contact angle of 151.2$^\circ$ $\pm$ 1.9$^\circ$. The other superhydrophobic surface was created by firstly coating PDMS on a silicon wafer, and then hydrophobic silica powder was distributed on the substrate surface and fixed by PDMS evenly through vibration. After heating and solidifying, the surface achieved a contact angle of 153.4$^\circ$ $\pm$ 2.3$^\circ$. According to the characteristics of their surface structures (see Figures \ref{fig:01}(b, c)), the two types of surfaces were named single-scale nano-structured (SN) superhydrophobic surface and hierarchical-scale micro-nano-structured (HMN) superhydrophobic surface, respectively.

\section{Results and discussion}\label{sec:3}
\begin{figure}
  \centering
  \includegraphics[width=0.95\columnwidth]{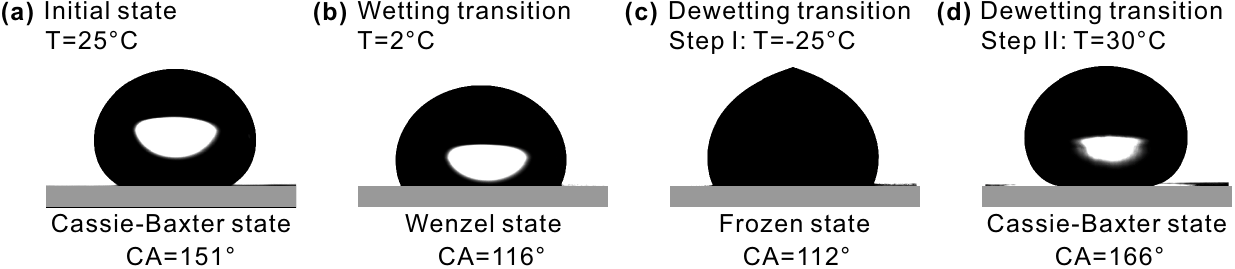}
  \caption{Wetting and dewetting transitions of a droplet on a superhydrophobic surface under the influence of temperature (Movie 1). Panels (a) and (b) show that the droplet undergoes a wetting transition due to the condensation on the superhydrophobic surface at a low temperature. Panels (c) and (d) show the freezing-melting strategy to recover the superhydrophobicity, which includes two steps: decreasing the surface temperature to freeze the droplet (i.e., Step I), and then increasing the surface temperature to melt the droplet (i.e., Step II). The heating temperature is 30 $^\circ$C, and the droplet volume is 37.1 µl.}\label{fig:02}
\end{figure}

In a cold and humid environment, condensation on solid surfaces is inevitable. Since superhydrophobic surfaces often have complex micro/nano-structures, if condensate droplets appear between these structures and fill the gaps, it is difficult to form airbags when droplets land on the surfaces \cite{kreder16, singh20}. Hence, the wetting transition under the condensation condition can easily occur, after which it is difficult for the substrate to recover the superhydrophobic property due to the energy barrier. Here, we find a strategy that can make droplets achieve a dewetting transition from the Wenzel state to the Cassie-Baxter state on the superhydrophobic surface. At the room temperature of 25 $^\circ$C, a droplet deposited on the superhydrophobic surface is in the Cassie Baxter state with a static contact angle of 151.2$^\circ$, as shown in Figure \ref{fig:02}(a). However, when the substrate temperature decreases below the dew point, a droplet deposited on this superhydrophobic surface is in the Wenzel state (i.e., the static contact angle is 116$^\circ$ at 2 $^\circ$C), as shown in Figure \ref{fig:02}(b). Even though the surface is still hydrophobic, the remarkable reduction in the contact angle indicates a wetting transition from the Cassie-Bxter state to the Wenzel state. This occurs because of the condensation on the surface, which will be explained later in \hyperref[sec:3.1]{Section 3.1}. Our strategy to recover the superhydrophobicity includes two steps. We first decrease the substrate temperature to -25 $^\circ$C to freeze the droplet (i.e., Step I as shown in Figure \ref{fig:02}(c)), and then we increase the substrate temperature to 30 $^\circ$C to melt the droplet (i.e., Step II as shown in Figure \ref{fig:02}(d)). Here, the substrate temperature is set at a temperature slightly higher than the room temperature, which can simplify the temperature control by using a heating device and fine-tuning the input voltage through an adjustable power supply. After the freezing-melting process, the static contact angle can recover to 166$^\circ$, i.e., a superhydrophobic state. It implies that a dewetting transition from the Wenzel to Cassie-Baxter state has occurred. To understand the mechanism of the phenomenon occurring under the above experimental conditions, the whole process from the wetting transition to the dewetting transition is analyzed in this study.

\subsection{Wetting transition}\label{sec:3.1}
\subsubsection{Condensation on the superhydrophobic surfaces}\label{sec:3.1.1}
The wetting transition can be explained by the condensation on the superhydrophobic surface at reduced temperatures, in which the condensate droplets form between the micro/nano-structures of the superhydrophobic surface, and inhibit the trapping of air when the droplets are deposited on the surface. Due to the diversity of micro/nano-structures on superhydrophobic surfaces, condensation may be different and then has different effects on the static contact states and the dynamic impact processes for millimeter droplets deposited on them. Therefore, we first characterize the condensation on superhydrophobic surfaces and analyze its influence on the wetting transition (from the Wenzel state to the Cassie-Baxter state) including static contact states and droplet impact processes.

\begin{figure}
  \centering
  \includegraphics[width=0.9\columnwidth]{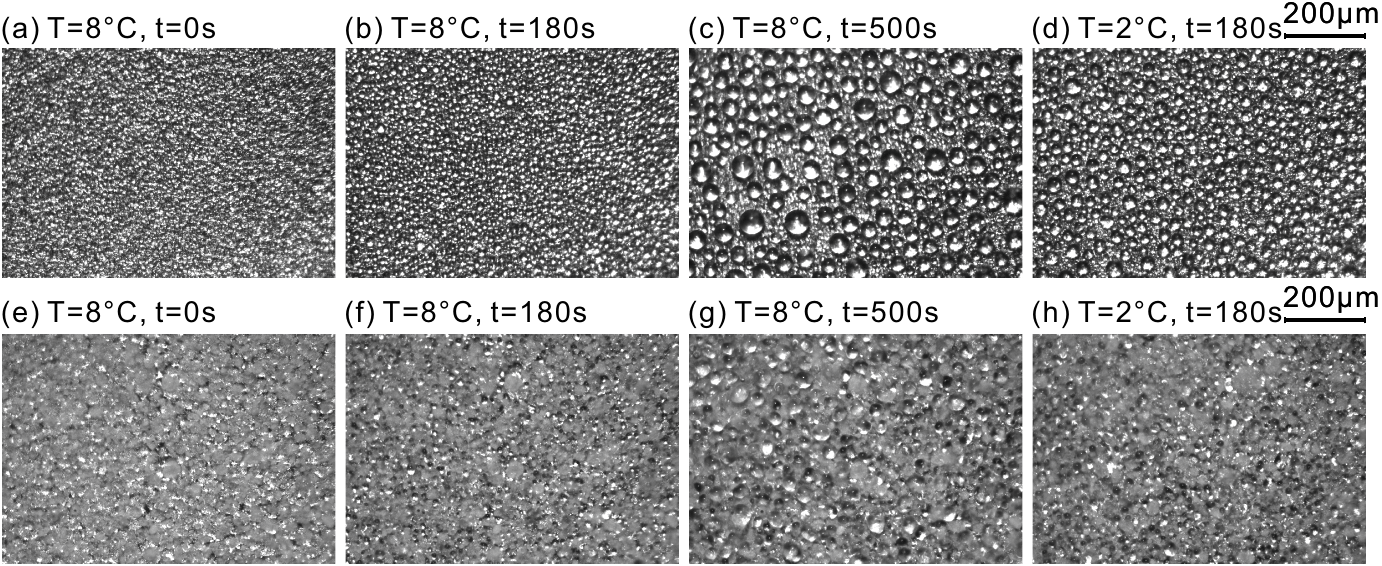}
  \caption{Condensation on (a) SN superhydrophobic surface and (b) HMN superhydrophobic surface: (a-c, e-g) image sequences showing the condensation condition over time at the substrate temperature of 8 $^\circ$C; (d, h) images showing the condensation state at the substrate temperature of 2 $^\circ$C and the cooling time of 180 s, for comparison with that in panels (b) and (f). To avoid the influence of observation position on the experimental results, the field of view of the camera is constrained within a region ($5 \times 5$ cm$^2$) at the center of the superhydrophobic surface ($40 \times 40$ cm$^2$).}\label{fig:03}
\end{figure}

In the experiment, the condensation processes on the superhydrophobic surfaces were recorded to get the distribution and growth characteristics of condensate droplets. The results of condensation on the superhydrophobic surfaces are shown in Figure \ref{fig:03}. When the ambient temperature around the substrate is lower than the dew point, moisture in the supersaturated air condenses on the surface, forming small droplets. Because the complex micro/nano-structures of the superhydrophobic surface could provide many nucleation sites, condensed droplets are more likely to appear within the interval of these micro/nano-structures, especially in the early stage of condensation. With the growth and coalescence of the condensate droplets (e.g., the condensation in Figures \ref{fig:03}(a-c)), the superhydrophobic surface changed markedly compared with the original state. Moreover, a lower temperature could accelerate the condensation process, and significantly larger droplets are condensed on the surface at the substrate temperature of 2 $^\circ$C (see Figure \ref{fig:03}(d)) than at the substrate temperature of 8 $^\circ$C (see Figure \ref{fig:03}(b)). In addition, the difference in the condensation process between the two types of superhydrophobic surfaces is also apparent. Only several small droplets can be found on the HMN surface (i.e., in Figure \ref{fig:03}(g)), which is in contrast with the condition on the SN surface (i.e., in Figure \ref{fig:03}(c)). The condensation on the superhydrophobic surfaces can cause a change in their superhydrophobicity, which will be discussed in Sections \hyperref[sec:3.1.2]{3.1.2} and \hyperref[sec:3.1.3]{3.1.3}.

\subsubsection{Influence of condensation on the static contact states}\label{sec:3.1.2}
\begin{figure}
  \centering
  \includegraphics[width=\columnwidth]{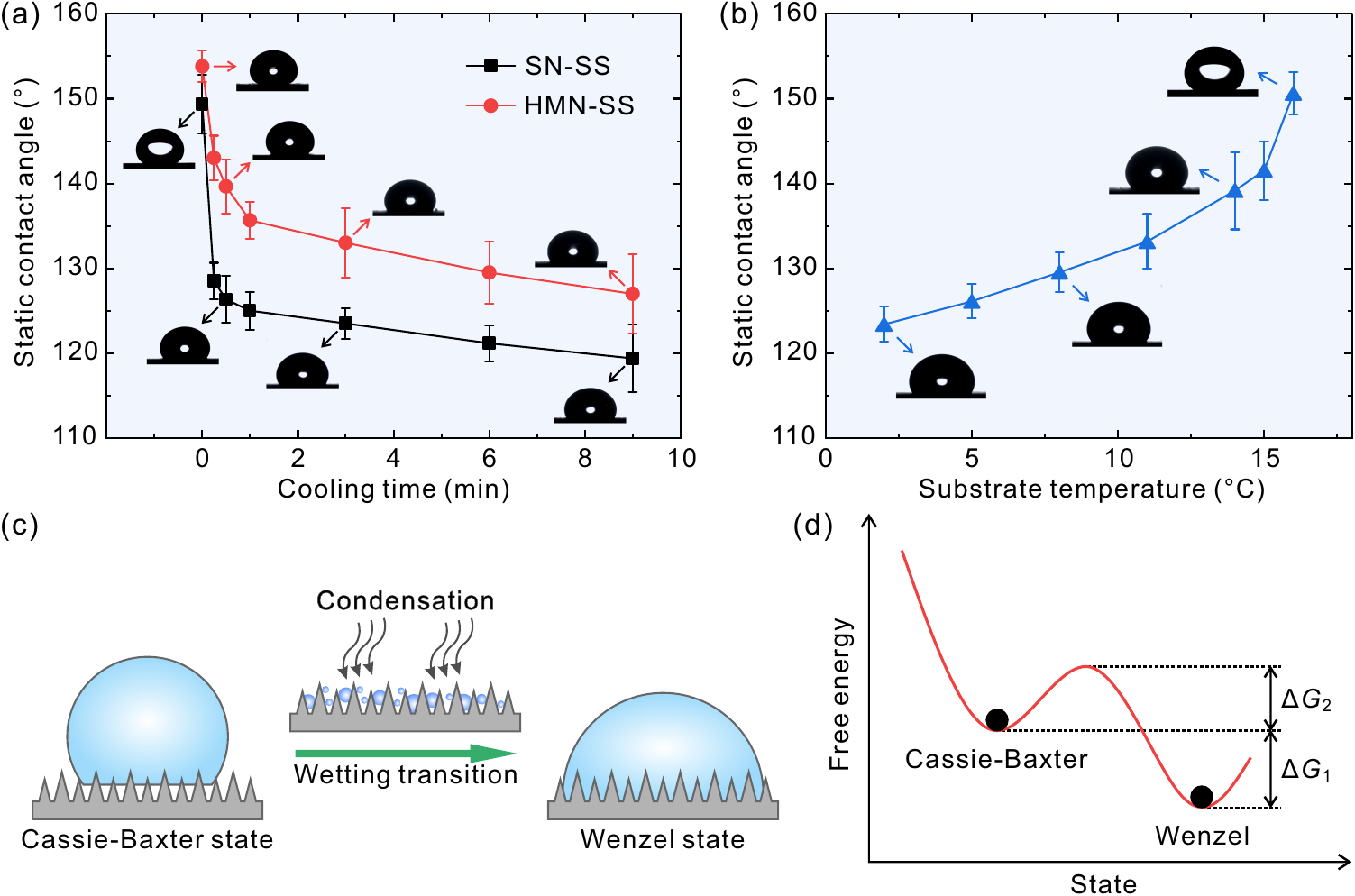}
  \caption{Static contact angles on the SN and HMN superhydrophobic surfaces under different experimental conditions, and wetting transitions for droplets on the superhydrophobic surface. (a) Variation of the static contact angle over the cooling time at the substrate temperature of 2 $^\circ$C. (b) Variation of the static contact angle against the substrate temperature for the SN superhydrophobic surface at the cooling time of 180 s. (c) Schematic illustration of wetting transition due to condensate droplets within the micro/nano-structures of the surface. (d) Variation of the free angle for different contact states of droplets on a superhydrophobic surface.}\label{fig:04}
\end{figure}

If the superhydrophobic surface is covered with many condensate droplets (whether in the gap of the microstructure or on the superhydrophobic surface), the static contact state of a millimeter droplet placed on this kind of surface may change and even lose its superhydrophobicity in severe cases. As shown in Figures \ref{fig:04}(a, b), the static contact angle decreases significantly even if there is only a small amount of condensate droplets on the SN and HMN surfaces (even after a short cooling time as shown in Figure \ref{fig:04}(a) or at a relatively higher substrate temperature as shown in Figure \ref{fig:04}(b)). Therefore, it is challenging to avoid the wetting transition and maintain the superhydrophobicity in cold environments.

Superhydrophobicity is usually achieved by maintaining the stability of the air layer beneath the droplet. However, condensate droplets can fill up the gaps of micro/nano-structures and force the air out, as illustrated in Figure \ref{fig:04}(c). Therefore, the wetting transition due to the condensation in a cold environment is a common phenomenon. Under the effect of condensate droplets, the contact state changes from the Cassie-Baxter state to the Wenzel state (as illustrated by the red curve in Figure \ref{fig:04}(d)), which is because the total free energy in the wetting state (i.e., the Wenzel state) is lower than that in the nonwetting state (i.e., the Cassie-Baxter state). According to previous theoretical analysis of the contact state transition \cite{bormashenko15, murakami14, ren14}, if the droplet recovers the nonwetting state, it needs to overcome the difference in the Gibbs free energy between the states before and after the transition (i.e., $\Delta G _1$). Additional free energy is also needed to overcome the energy barrier in the change process (i.e., $\Delta G _2$). In general, it is relatively easy for the wetting process to occur, achieving a wetting transition from the Cassie-Baxter state to the Wenzel state, during which the energy needed is only $\Delta G _2$. However, it is difficult for the dewetting process to occur in the opposite direction, during which the energy barrier is $\Delta G _1 + \Delta G _2$. Another evidence for the difficult dewetting transition is that the contact area of the droplet in the Wenzel state is larger than that in the Cassie-Baxter state, which greatly increases the surface adhesion. If there is sufficient free energy input, it is possible for the dewetting process to occur, thus achieving a dewetting transition.

\subsubsection{Influence of condensation on droplet impact processes}\label{sec:3.1.3}
\begin{figure}
  \centering
  \includegraphics[width=\columnwidth]{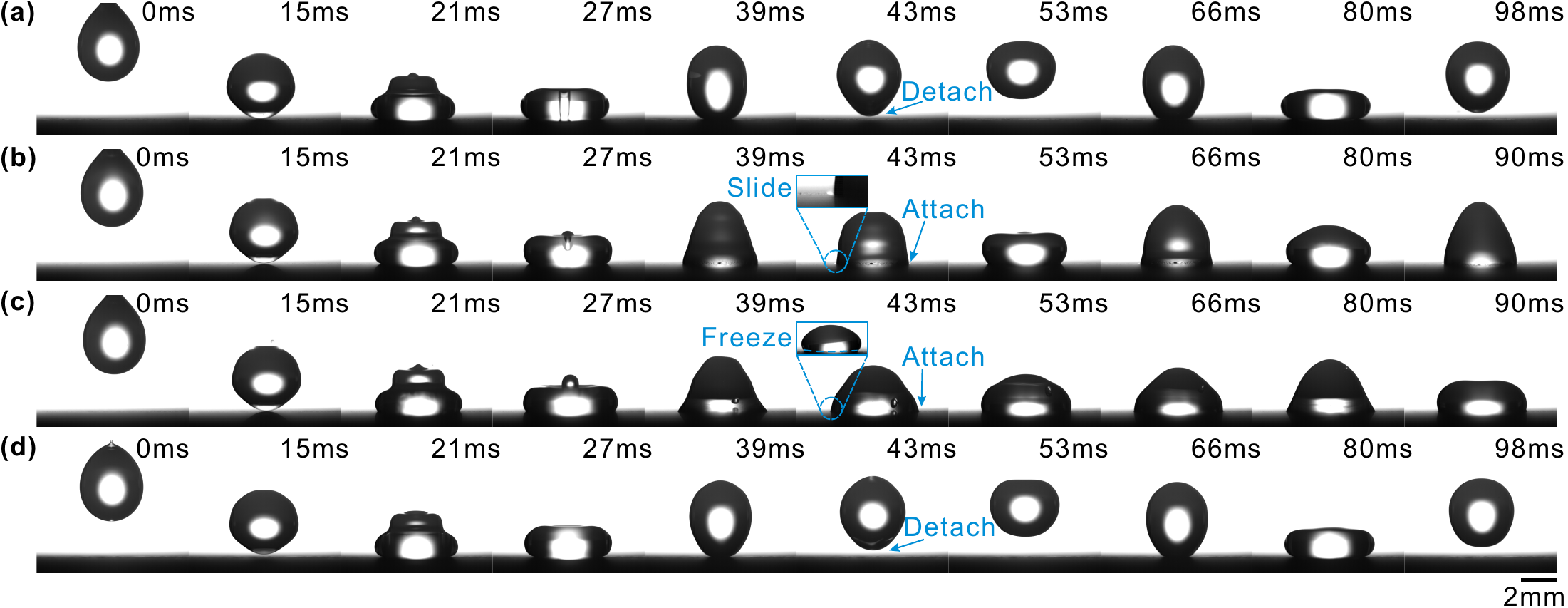}
  \caption{Image sequences showing the impact processes of droplets on the SN and HMN superhydrophobic surfaces with different degrees of condensation: (a) for the SN substrate at 30 $^\circ$C (i.e., without cooling, Movie 2); (b) for the SN substrate with a temperature of 2 $^\circ$C for condensation (Movie 3); (c) for the SN substrate with a temperature of -25 $^\circ$C for condensation and freezing (Movie 4); (d) for the HMN substrate with a temperature of 2 $^\circ$C for condensation.}\label{fig:05}
\end{figure}

\begin{figure}
  \centering
  \includegraphics[width=0.5\columnwidth]{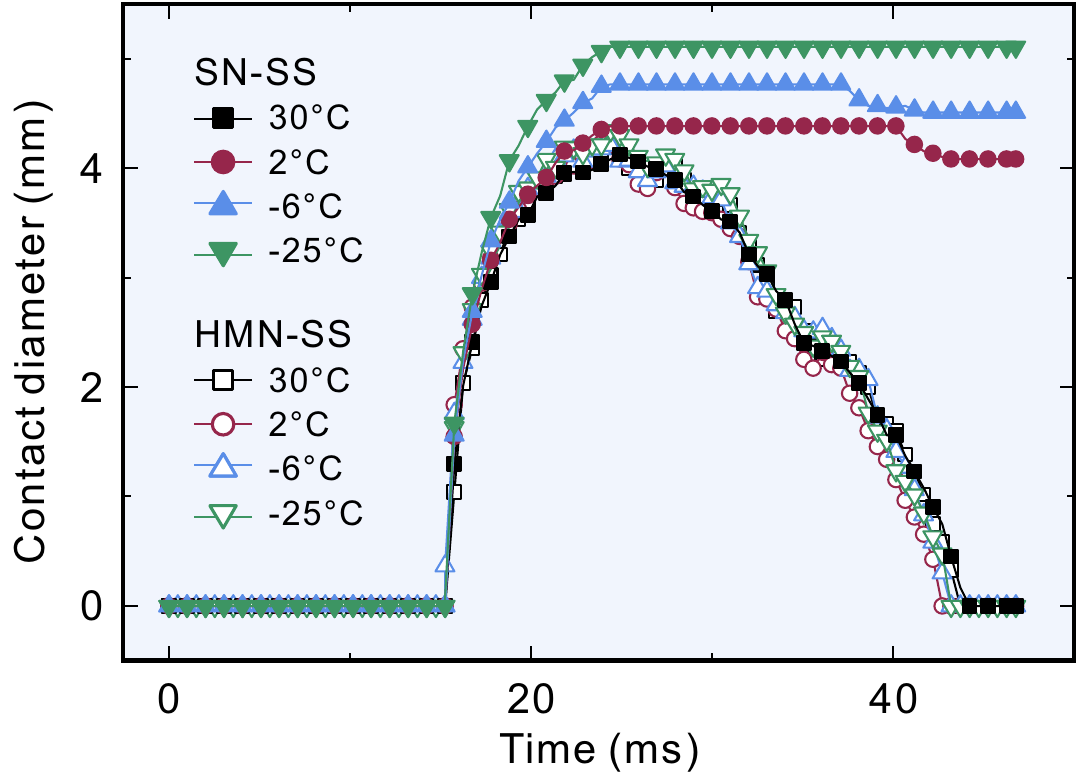}
  \caption{Contact diameters during droplet impact on the superhydrophobic substrates with different substrate temperatures.}\label{fig:06}
\end{figure}

Due to the variation in the surface condition under the influence of condensation, the surface adhesion force in this condition can be greater and hinder the movement of droplets than in dry conditions. The impact processes of droplets on the superhydrophobic substrates with different degrees of condensation are shown in Figure \ref{fig:05}. Due to the similar static contact angles on the SN superhydrophobic surface (CA = 151.2$^\circ$) and HMN superhydrophobic surface (CA = 153.4$^\circ$), their impact processes are similar at the substrate temperature of 30 $^\circ$C. They both have the typical feature of droplets bouncing on superhydrophobic surfaces. However, with the decrease in the substrate temperature, the fast-spreading process during the impact changes significantly and the droplet cannot detach from the surface at all times due to the loss of surface superhydrophobicity and the increase in surface adhesion under the influence of condensation, as shown in Figure \ref{fig:05}(b). Note that, at $t = 43$ ms, the droplet shrinks a little, which indicates that the loss of superhydrophobicity is a gradually changing process. A more extreme phenomenon is that the bottom of the droplet has frozen when the impact droplet reaches the maximum spreading (as shown in $t = 43$ ms of Figure \ref{fig:05}(c)), which prevents the subsequent contraction of the droplet and the reduction in its contact diameter (as shown quantitatively in Figure \ref{fig:06}). As for the impact process on the HMN surface, there is no significant change in the droplet deformation as the temperature decreases from 30 $^\circ$C to -25 $^\circ$C. Here, the image sequence of droplet impact on the substrate with the temperature of 2 $^\circ$C is shown in Figure \ref{fig:05}(d), which is similar to that in Figure \ref{fig:05}(a). This independence of temperature is consistent with the results of the surface condensation conditions shown in Figure \ref{fig:03}(e-g): to a certain extent, the HMN superhydrophobic surface can maintain its superhydrophobicity even in a cold environment.

To quantitatively characterize the effect of the condensation on the impact process, the contact diameter during the impact process on different superhydrophobic surfaces and different substrate temperatures is measured, see Figure \ref{fig:06}. On the SN surface, the maximum spreading diameter increases as the substrate temperature decreases from 30 $^\circ$C to -25 $^\circ$C, resulting in the deterioration of superhydrophobicity. In addition, at 30 $^\circ$C, the droplet, after spreading to the maximum contact area, immediately shrinks and moves upward. In contrast, at 2 $^\circ$C, -6$^\circ$C, and -25 $^\circ$C, the droplet could maintain its maximum contact area for a while even under the influence of vibrant oscillation. Finally, there is always a very small decrease in the contact diameter due to the contraction caused by repeated oscillations of the droplet, except at -25 $^\circ$C, which has made the bottom of the droplet frozen in the spreading stage. For the HMN surface, in comparison, the changes in the contact diameter at 30 $^\circ$C, 2 $^\circ$C, -6$^\circ$C and -25 $^\circ$C are similar to the case on the SN surface at 30 $^\circ$C, indicating that the anti-condensation property of HMN surface plays an important role in maintaining the surface superhydrophobicity.

\subsection{Dewetting transition}\label{sec:3.2}
\begin{figure}
  \centering
  \includegraphics[width=\columnwidth]{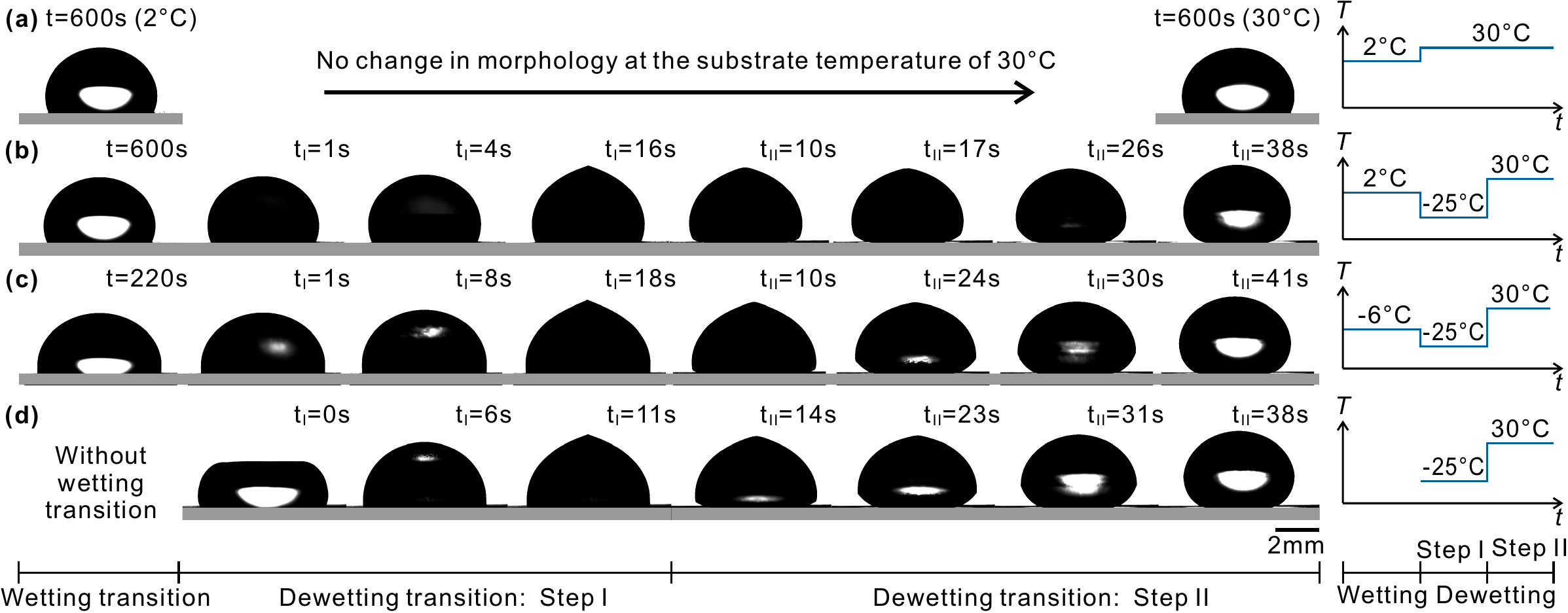}
  \caption{Dewetting transitions on the SN superhydrophobic surface under different temperature sequences. (a) The droplet is deposited on the substrate which has been cooled to 2 $^\circ$C for 600 s. Then the substrate temperature is increased to 30 $^\circ$C and maintained for 600 s. (b) The droplet is deposited on the substrate which has been cooled to 2 $^\circ$C for 600 s. Then the substrate temperature is reduced to -25 $^\circ$C to freeze the droplet. After the droplet has completely frozen, the substrate temperature is increased to 30 $^\circ$C to make the frozen droplet melt. (c) The whole process of the temperature setup is the same as that in panel (b) except that the substrate temperature is -6 $^\circ$C during the wetting transition. (d) The droplet is deposited on the substrate which has been cooled to -25 $^\circ$C for 60 s (Movie 5). The droplet freezes immediately upon the deposition. After the droplet freezes completely, the substrate temperature is increased to 30 $^\circ$C to make the frozen droplet melt. In panels (b-d), in the freezing step, the moment when the nucleation starts is defined as $t_\text{I} = 0$, and in the melting step, the moment when the melting starts is defined as $t_\text{II} = 0$. The droplet volume is 37.1 µl.}\label{fig:07}
\end{figure}

We then explore the method and mechanism for achieving the dewetting transition (i.e., from the Wenzel state to the Cassie-Baxter state) by controlling the substrate temperature. The results presented in \hyperref[sec:3.1.1]{Section 3.1.1} have shown that droplets change to the Wezel state at reduced temperature because of the condensation. Here, if we directly increase the substrate temperature to a warmer condition (above the dew-point temperature, e.g., 30 $^\circ$C), the wetting state does not change, as shown in Figure \ref{fig:07}(a). As mentioned at the beginning of \hyperref[sec:3]{Section 3}, our strategy for dewetting transition relies on a more complex process, i.e., two steps including the freezing and then the melting of the droplet. As shown in Figure \ref{fig:07}(b), the droplet is deposited on the SN surface which is in the same state as that in Figure \ref{fig:07}(a). The substrate is first cooled down to -25 $^\circ$C (i.e., Step I) to make the droplet freeze, then it is heated to 30 $^\circ$C (i.e., Step II) to make the droplet melt. In the melting process of the frozen droplet with a Wenzel-like morphology, there are obvious changes in the droplet morphology that the contact area decreases and the contact angle increases to 166$^\circ$. This indicates that the SN surface has recovered its superhydrophobicity. The freezing-melting strategy is also tested for the substrate with a lower condensation temperature (i.e., -6 $^\circ$C), as shown in Figure \ref{fig:07}(c). The droplet at -6 $^\circ$C is in the supercooled state and experiences a more serious wetting transition than that at 2 $^\circ$C. Therefore, in comparison, the static contact angle further decreases from 116$^\circ$ at 2 $^\circ$C to 102$^\circ$ at -6 $^\circ$C. Although the morphologies of the liquid droplet and its frozen droplet are slightly different between these two cases, their contact angles after the melting are similar, as shown in $t_\text{II} = 38$ s in Figure \ref{fig:07}(b) and $t_\text{II} = 41$ s in Figure \ref{fig:07}(c). We also conduct an experiment in which a droplet is deposited on the SN surface with a very low temperature of -25 $^\circ$C (as shown in Figure \ref{fig:07}(d)). The droplet freezes immediately upon contact with the substrate and finally exhibits a Wenzel-like morphology. As the substrate temperature increases to 30 $^\circ$C, the melted droplet can still recover to the Cassie-Baxter state. In general, whether the droplet is deposited in a non-supercooled state, or a supercooled state, or immediately freezes upon contact with the SN surface, it can always achieve a dewetting transition through the freezing-melting process.

\begin{figure}
  \centering
  \includegraphics[width=\columnwidth]{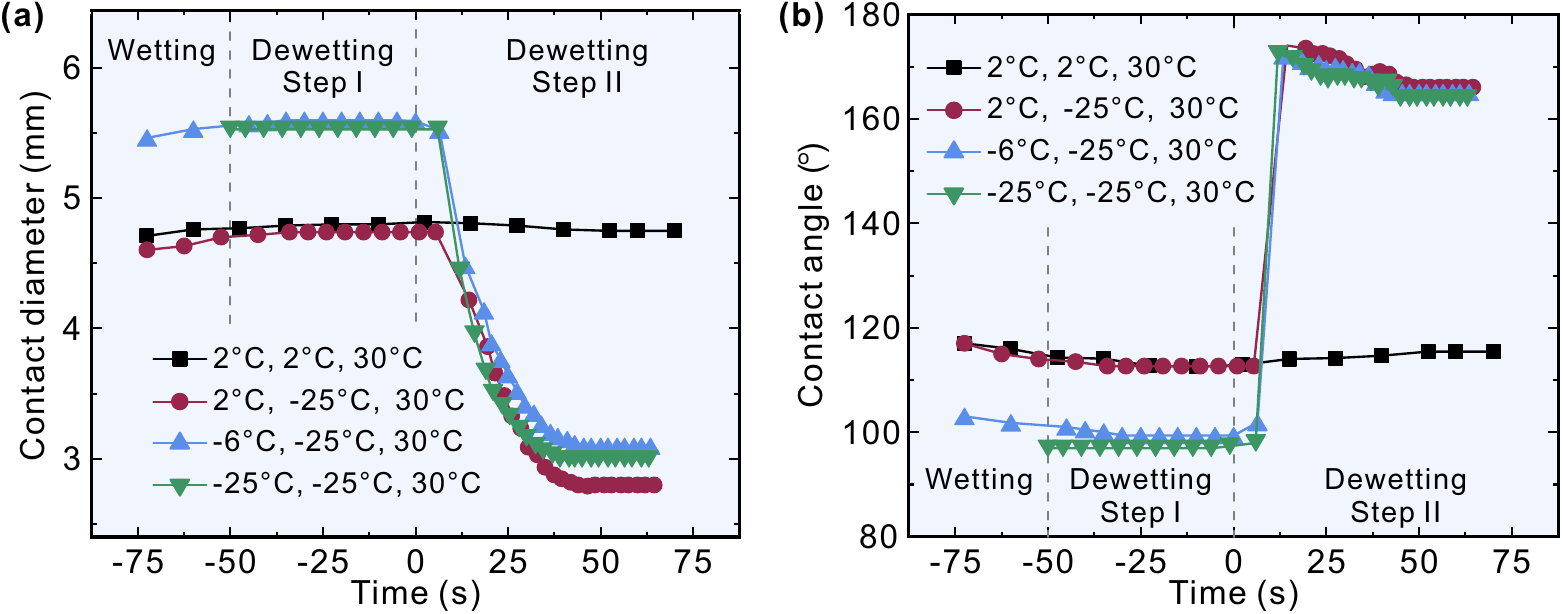}
  \caption{(a) Contact diameter and (b) contact angle of droplets on the SN superhydrophobic surface during the dewetting transition under different temperature conditions corresponding to Figure \ref{fig:07}. The process can be divided into three stages, namely, wetting, dewetting: Step I (i.e., freezing stage), and dewetting: step II (i.e., melting stage). The substrate temperatures in the three stages are provided in the legends of the figures sequentially. In the plots, the starting time of the freezing stage is set as $t= -50$ s, and the starting time of the melting process is set as $t = 0$.
}\label{fig:08}
\end{figure}

The contact diameter and contact angle of the droplet on the SN superhydrophobic surface during the dewetting transition are analyzed based on digital image processing, and the results are shown in Figure \ref{fig:08}. After the freezing-melting processes, all the droplets can achieve a dewetting transition. In the early stage of melting, the contact diameter has a notable decrease and the contact angle has a significant increase, indicating that the dewetting transition mainly occurs in this stage. The droplet returns to the Cassie-Baxter state before the droplet completely melts. These results mean that the freezing-melting process has an excellent dewetting capability for droplets on the SN superhydrophobic surface.

\begin{figure}
  \centering
  \includegraphics[width=\columnwidth]{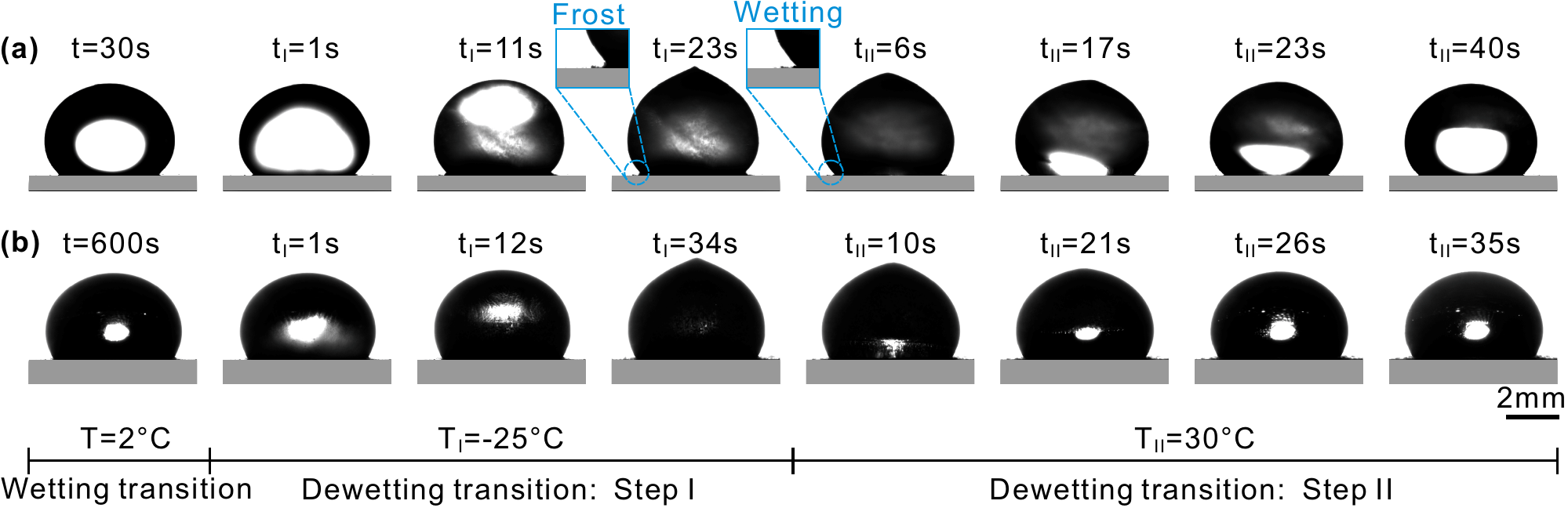}
  \caption{Images sequences showing that droplets on the HMN superhydrophobic surface fail to achieve the dewetting transition through the freezing-melting process: (a) the cooling time for condensation before the wetting transition is 30 s; (b) the cooling time for condensation before the wetting transition is 600 s. The condensation in (a) and (b) have different degrees, but the freezing-melting strategy for the attempt of dewetting transition in (a) and (b) is the same as that in Figure \ref{fig:07}(b). The droplet volume is 37.1 µl.}\label{fig:09}
\end{figure}

For comparison, we also performed experiments with the HMN superhydrophobic surface under the same freezing-melting strategy as that in Figure \ref{fig:07}(b). Regardless of what degree of wetting transition (at different substrate temperatures from 30 $^\circ$C to -25 $^\circ$C) or what freezing-melting strategy (either decreasing the substrate temperature to -25 $^\circ$C after the deposition, or directly reducing the substrate temperature to -25 $^\circ$C before the deposition), the contact angle of the droplet after the melting has no increase comparing with the state after the wetting transition (as shown in Figure \ref{fig:09}). We even found that the frost formed near the air-water-solid three-phase contact line may melt and merge with the melting droplet, which further increases the wetting of the droplet, as shown at $t_\text{I} = 23$ s and $t_\text{II} = 6$ s in Figure \ref{fig:09}(a), respectively.

\begin{figure}
  \centering
  \includegraphics[width=0.9\columnwidth]{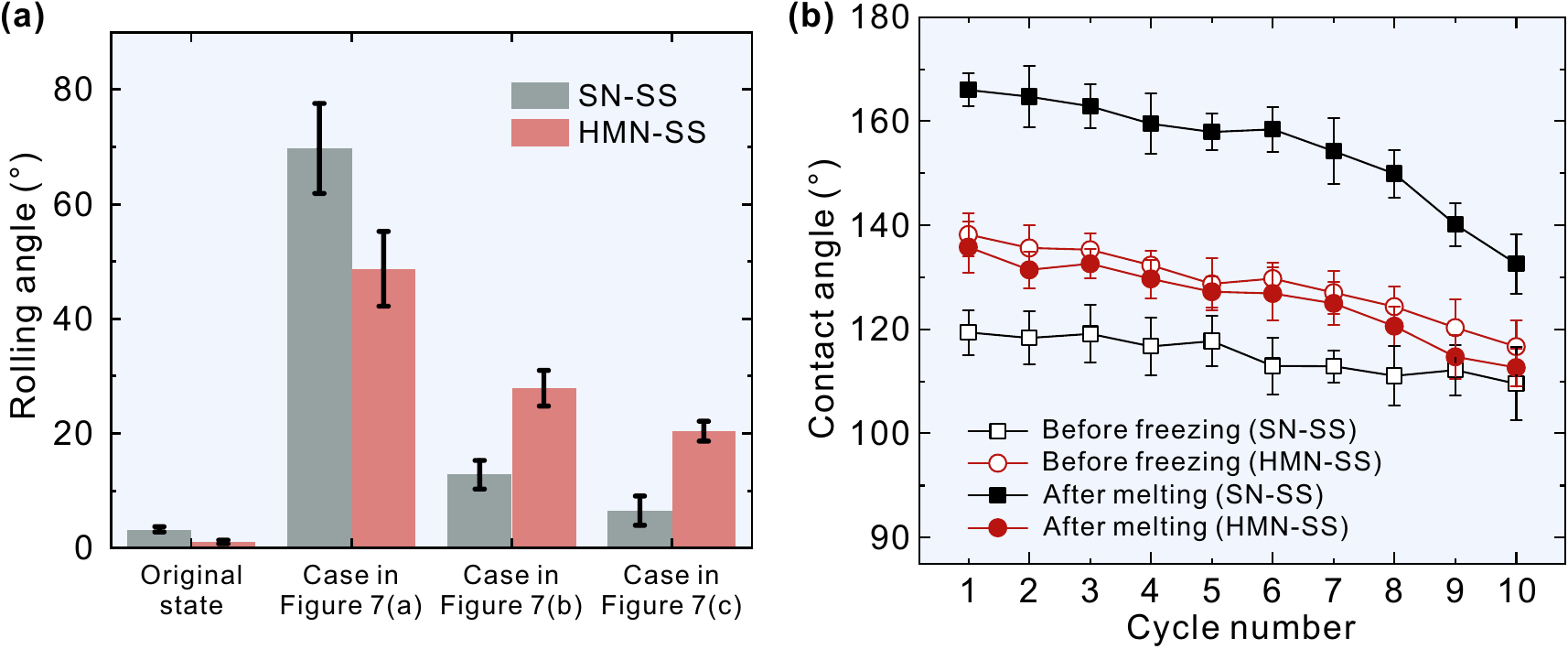}
  \caption{Change of the surface superhydrophobicity. (a) Rolling angle of droplets under different temperature sequences. As for the original static contact angles of these two superhydrophobic surfaces, droplets were deposited on the substrates in a laboratory environment with a temperature of 23--25 $^\circ$C and a humidity of 40–50\%. The other three cases correspond to the processes in Figures \ref{fig:07}(a, b, d). The rolling angles were obtained by changing the tilt angle of substrates until the droplets were about to roll. (b) Deterioration in the contact angle after the wetting and dewetting transitions during cycled freezing-melting processes.}
  \label{fig:10}
\end{figure}

To further check the surface superhydrophobicity, we measure the rolling angle before the condensation and after the melting. A small rolling angle is a typical feature of superhydrophobic surfaces. Due to the existence of contact angle hysteresis, the movement of droplets on surfaces is limited, especially after the wetting phenomenon occurs, which makes it more difficult to remove these droplets through simple methods such as surface tilting \cite{wen14}, air flowing \cite{yeganehdoust21}, or droplet coalescence \cite{farhangi12}. Therefore, the variation of the rolling angle under the freezing-melting strategy can reflect the change of interfacial energy during the wetting and dewetting processes. The variation of the rolling angle of superhydrophobic surfaces is shown in Figure \ref{fig:10}(a). In the experiment, the rolling angle was measured by adjusting the tilt angle of an angle swing device below the superhydrophobic surface at a rate of 0.5$^\circ$ in every 30 s and recorded by the camera from the side view. When the droplet was about to roll off, the value of the angle swing device was determined as the rolling angle. Before the condensation, the rolling angles of the SN and HMN superhydrophobic surfaces are very small, i.e., 3.2$^\circ$ and 1.1$^\circ$, respectively. The small rolling angles indicate that droplets deposited on the surfaces can easily roll away and these surfaces have good superhydrophobicity. After condensation, the rolling angle on both superhydrophobic surfaces increases significantly, making it difficult for droplets to detach. For the cases in Figure \ref{fig:07}(b,c), the increments in the rolling angle after the freezing-melting process for dewetting are much smaller than that in Figure \ref{fig:07}(a), even though the rolling angle also increases compared with that before the condensation. This result indicates that the freezing-melting process alleviates the deterioration of the surface superhydrophobicity.

To explore the repeatability of the dewetting characteristic on the SN superhydrophobic surface during repeated freezing-melting cycles, the contact angle was measured after the wetting and dewetting transitions when the droplet state stabilized, as shown in Figure \ref{fig:10}(b). In the experiment, the deterioration was tested in the same position of the same superhydrophobic surface. When the contact angle of a droplet before freezing and after melting were recorded, the droplet was removed and a new droplet was placed in the same position for a new freezing-melting process. In each cycle, the static contact angle has a significant increase after the freezing-melting process, which indicates that the droplet can effectively achieve the dewetting transition. As for the HMN superhydrophobic surface, there is a small reduction in the contact angle after the freezing-melting process due to merging with nearby droplets that are generated by the melting of frost (as shown in Figure \ref{fig:09}(a)). Therefore, the dewetting transition on the HMN superhydrophobic surface is not achieved. Another obvious characteristic in Figure \ref{fig:10}(b) is that the contact angle after melting gradually decreases. It can be attributed to the fragility of the micro/nano-structures on superhydrophobic surfaces, which will be further discussed in the next section.

\subsection{Mechanism of the dewetting transition}\label{sec:3.3}
\begin{figure}
  \centering
  \includegraphics[width=0.8\columnwidth]{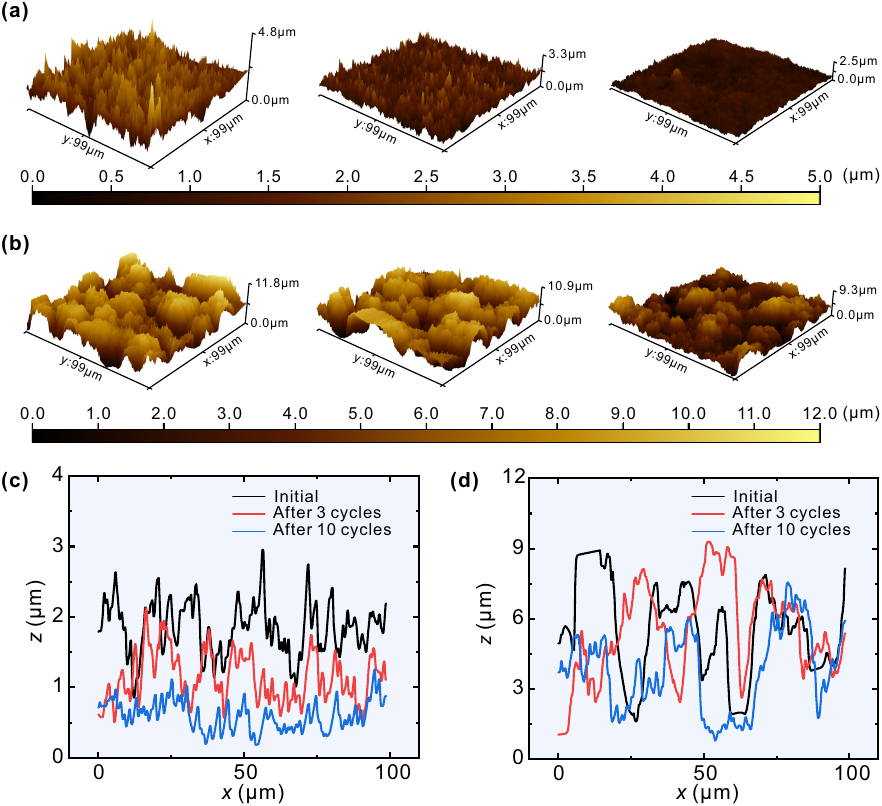}
  \caption{AFM images and surface profiles of the micro/nano-structures on superhydrophobic substrates during freezing-melting cycles: (a, b) AFM images initially and after 3 and 10 cycles on the SN (a) and HMN (b) superhydrophobic surfaces, respectively; (c, d) surface profiles of the SN (c) and HMN (d) superhydrophobic substrates along the $y$-axis, respectively.}\label{fig:11}
\end{figure}

To analyze the mechanism of the dewetting transition, the differences in the micro/nano-structures of the superhydrophobic surfaces are explored. The micro/nano-structures of the superhydrophobic surfaces are obtained by atomic force microscope (AFM). As shown in Figure \ref{fig:11}, there is a significant difference between the SN and HMN superhydrophobic substrates. The SN surface has a slender needle-shaped structure, while the HMN surface has a robust spherical-shaped structure. Hence, it is obvious that the latter has better geometric strength. In addition, after a certain freezing-melting cycle, both superhydrophobic surfaces have undergone a certain degree of structural damage. Many needle-shaped structures on the SN surface have fallen off, as shown in Figure \ref{fig:11}(a), which is in contrast to the change on the HMN surface with only a small number of spherical-shaped structures disappearing, as shown in Figure \ref{fig:11}(b). The changes in the micro/nano-structures can also be proved by the profiles of the SN and HMN surfaces, as shown in Figures \ref{fig:11}(c, d). These results indicate that the processes of the wetting and dewetting transitions in the experiment have made effects on the structures of the superhydrophobic surfaces with certain damage on the SN surface and negligible damage on the HMN surface.

\begin{figure}
  \centering
  \includegraphics[width=0.85\columnwidth]{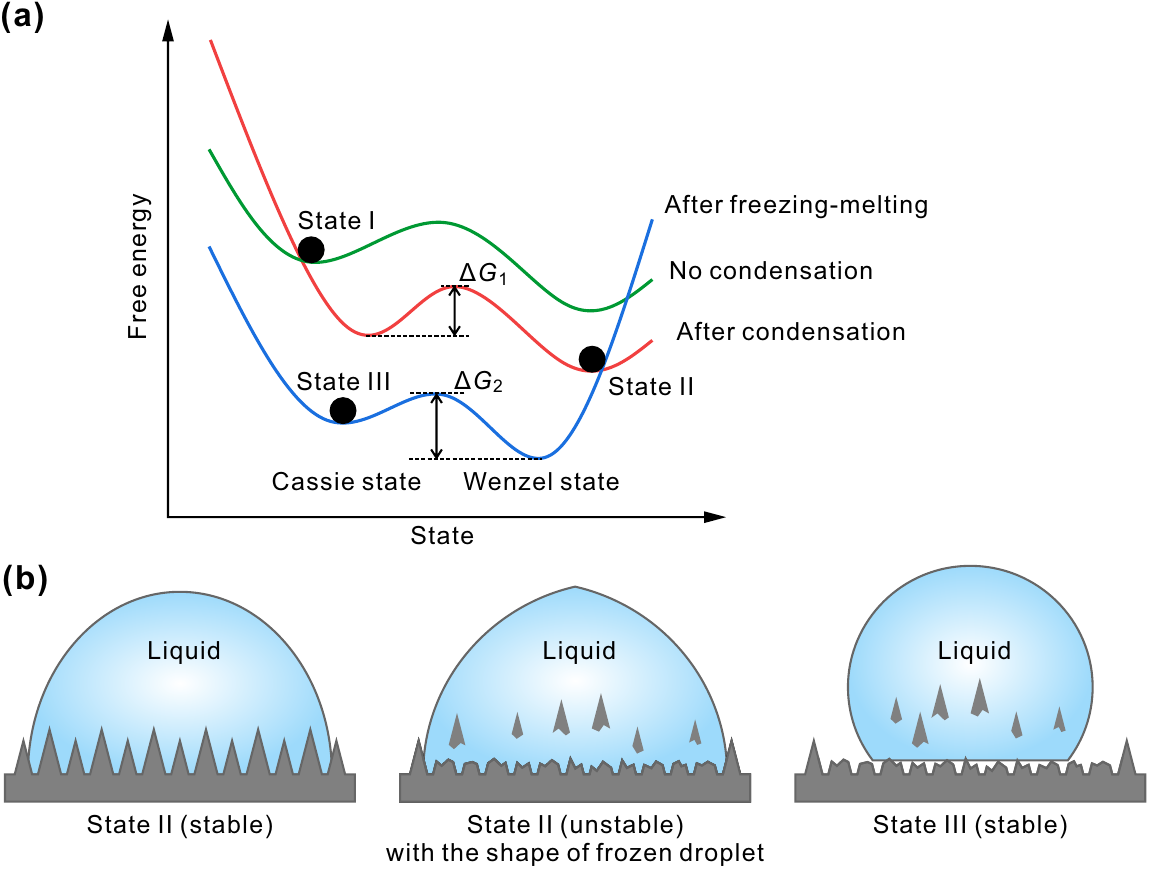}
  \caption{(a) Energy states of the droplet during the wetting and dewetting transitions. (b) Schematic illustration of the dewetting transition.}\label{fig:12}
\end{figure}

Next, the changes in the contact state on the SN superhydrophobic surface during the wetting and dewetting transitions are analyzed from the perspective of the energy state. As discussed in Figure \ref{fig:04}(d), droplets on superhydrophobic surfaces have two relatively stable states, the Cassie-Baxter state and the Wenzel state, where the free energy of the system is at a local minimum (as shown in Figure \ref{fig:12}(a)). For droplets on the SN superhydrophobic surface with the Cassie-Baxter state (i.e., State I on the green curve in Figure \ref{fig:12}(a)), the condensation on the substrate surface causes the disappearance of the airbag constrained between the micro/nano-structures, as discussed in \hyperref[sec:3.1.1]{Section 3.1.1}. Then, the contact state transforms into the Wenzel state (i.e., State II on the red curve in Figure \ref{fig:12}(a), which is a new curve of free energy due to the condensation on the SN superhydrophobic surface). After the droplet freezes, the connection between the micro/nano-structures and the substrate becomes very weak due to the compression of ice, and some of them are even damaged, which affects the curve of free energy (i.e., State II on the blue curve in Figure \ref{fig:12}(a)). Due to the change of the superhydrophobic surface, the contact state (i.e., State II on the red curve in Figure \ref{fig:12}(a)) is unstable if the droplet is in the liquid condition. Once the frozen droplet melts, the contact area will contract on this new surface under the effect of surface tension (as shown in Figure \ref{fig:12}(b)). During the dewetting transition, the contact state of the melting droplet gradually changes according to the schematic of the blue curve in Figure \ref{fig:12}(a). The released free energy overcomes the barrier from the Wenzel state (i.e., State II on the red curve) to the Cassie-Baxter state (i.e., State III on the blue curve). In general, the change of the SN superhydrophobic surface results in an unstable state for the melting droplet on this changed surface and can make the droplet overcome the barrier to achieve the dewetting transition with the help of surface tension (or the released free energy). This kind of dewetting transition is due to the change of the superhydrophobic surface, which is different from the use of external activations to achieve the dewetting transition, such as surface vibration \cite{boreyko09, lei14}, electric field \cite{he21, wikramanayake20}, magnetic field \cite{yang18, cheng12}, and droplet growth or coalescence \cite{cunjing15, zhang16}. All of these studies aim to change the contact state by directly acting on droplets, while our study alters the relationship between droplets and surfaces by acting on superhydrophobic surfaces through a freezing-melting process, which provides a new perspective for achieving the transition of the contact state.

In summary, two types of superhydrophobic surfaces with similar chemical composition and different micro/nano-structures are used in the experiment and the results on the dewetting transition are completely different. The comparison between these two superhydrophobic surfaces indicates that although nanoscale structures have low structural strength, this kind of superhydrophobic surface can promote the recovery of the superhydrophobicity under a certain freezing-melting strategy in the case of wetting. For superhydrophobic surfaces with high structural strength, the freezing and melting process cannot weaken the binding effect of these structures on the droplet (as shown in Figure \ref{fig:11}(b)), so the transition of the contact state is hard to achieve. Generally speaking, for superhydrophobic surfaces with relatively low mechanical strength but not completely destroyed during the process of droplet freezing, it is possible to achieve the dewetting transition during the process of droplet melting. In addition, regardless of whether the temperature is below the dew point but above the freezing point or below the freezing point (i.e., whether the droplet is in a liquid state or a solid state initially), the freezing-melting process is useful for the transition of contact state. This means that the freezing-melting strategy is applicable over a wide range of temperatures.

\begin{figure}
  \centering
  \includegraphics[width=0.95\columnwidth]{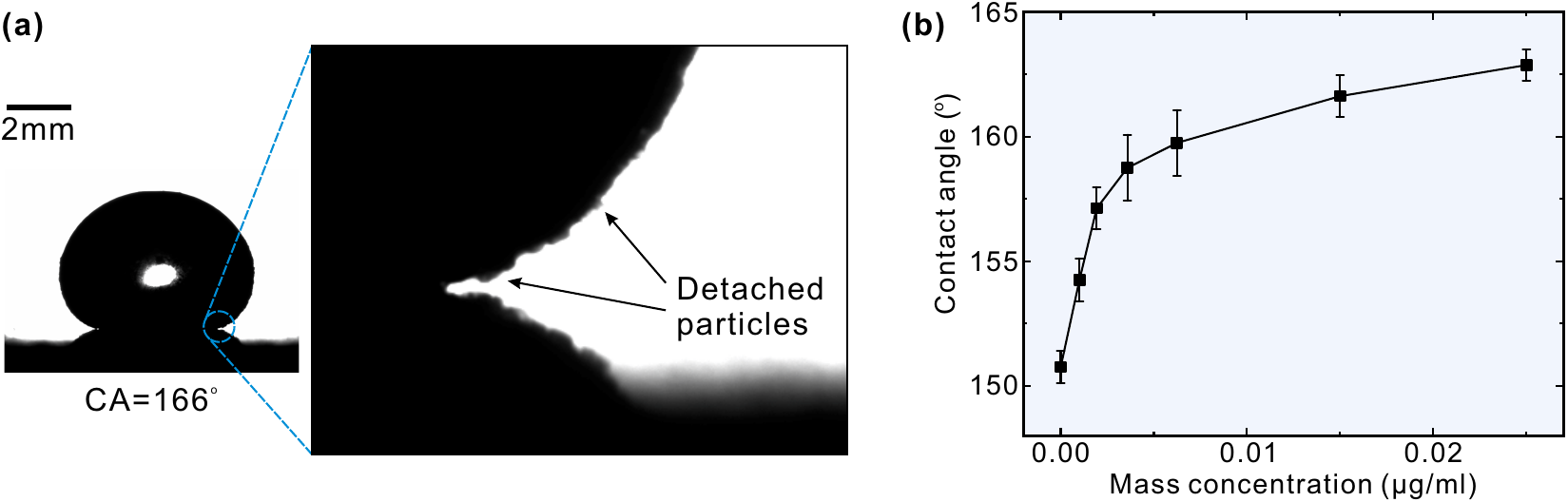}
  \caption{Effect of detached particles on the static contact angle of the droplet: (a) silhouette of the droplet after the freezing-melting cycle; (b) static contact angle of the droplet with different mass concentrations of monodispersed silica microspheres.}\label{fig:13}
\end{figure}

There is still something that needs further clarification regarding the surface damage during the freezing-melting strategy, i.e., the detachment of particles from the superhydrophobic surface. As shown in Figure \ref{fig:08}(b), the contact angle of all droplets on the SN superhydrophobic surface after the melting is slightly larger than the initial contact angle (i.e., 151.2$^\circ$ at the room temperature of 25 $^\circ$C). To explain why the contact angle increases after ice melting, the morphology of a droplet on the SN superhydrophobic surface after the freezing-melting cycle is shown in Figure \ref{fig:13}(a). The interface near the three-phase contact line of the droplet is very rough, which indicates that some solid particles are accumulated at the gas-liquid interface. Considering the damage to these surface nanostructures, we hypothesize that the accumulation of particles shedding from the SN superhydrophobic surface during the freezing-melting cycle increases the contact angle of the droplet. To confirm this hypothesis, we further explore the effect of particles on the static contact angle of the droplet. This is done by adding nanoparticles (monodispersed silica microspheres with an average diameter of 500 nm) to the droplet and then measuring the static contact angle at the room temperature of 25 $^\circ$C. As shown in Figure \ref{fig:13}(b), there is a significant increase in the droplet contact angle as the particle concentration increases, even with only a small amount of silica nanoparticles. This could be attributed to the interactions between particles or between particles and interfaces \cite{shao20, jiang16}.

\section{Conclusions}
Condensation can seriously affect the contact state of droplets on superhydrophobic surfaces, making them prone to undergo a wetting transition from the Cassie-Baxter state to the Wenzel state. In this study, two kinds of superhydrophobic surfaces (i.e., the SN and HMN superhydrophobic surfaces) are used to explore the condensation conditions. With the decrease of cooling temperature and increase in cooling time, the condensation on the surface becomes more severe. Comparing the two types of superhydrophobic surfaces, there are many condensate droplets on the SN surface, while only a few condensation droplets on the HMN surface under the same experiment conditions. The condensation affects the static contact angles of droplets deposited on these surfaces and inhibits the rebound of droplets after impact. To recover the surface superhydrophobicity, we propose a freezing-melting strategy to achieve the dewetting transition from the Wenzel state to the Cassie-Baxter state.

 The freezing-melting strategy requires freezing the droplet by reducing the substrate temperature and then melting the droplet by heating the substrate. The experiment results show that this method can be used over a wide temperature range from below dew-point temperature to below freezing-point temperature. The mechanism for the dewetting transition is analyzed by exploring the differences in the micro/nano-structures of the superhydrophobic surfaces and it is attributed to the unique structure and strength of the superhydrophobic surface.

 The freezing-melting process changes the energy states of the droplet on the surface, which makes the Wenzel state unstable, and then the droplet achieves a dewetting transition to a relatively stable Cassie-Baxter state. Since this strategy can be achieved easily by only adjusting the surface temperature, it has great application prospects in fields that require long-term maintenance of superhydrophobicity.

\begin{acknowledgement}
This work is supported by the National Natural Science Foundation of China (Grant Nos.\ 52176083, 51920105010, and 51921004).
\end{acknowledgement}

\begin{suppinfo}

The following files are available free of charge.
\begin{itemize}
  \item Movie 1: Video clip showing the wetting and dewetting transitions of a droplet on a SN substrate, corresponding to Figure \ref{fig:02}.
  \item Movie 2: Video clip showing the rebound of a droplet on a SN substrate at 30 $^\circ$C, corresponding to Figure \ref{fig:05}a.
  \item Movie 3: Video clip showing the deposition of a droplet on a SN substrate at -2 $^\circ$C, corresponding to Figure \ref{fig:05}b.  
  \item Movie 4: Video clip showing the freezing of a droplet on a SN substrate at -25 $^\circ$C, corresponding to Figure \ref{fig:05}c.    
  \item Movie 5: Video clip showing the dewetting transition of a droplet on a SN substrate, corresponding to Figure \ref{fig:07}d.
\end{itemize}

\end{suppinfo}

\bibliography{DewettingTransition}

\begin{tocentry}
\centering
\includegraphics{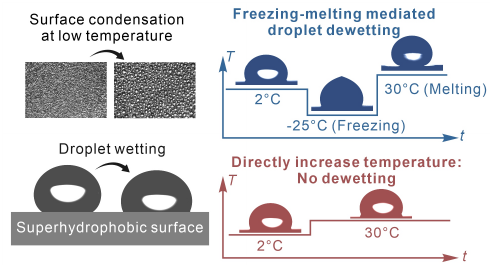}\\
\vspace{3mm}
\end{tocentry}
\newpage

\end{document}